\documentclass[12pt,preprint]{aastex}

\begin{document}

\title{Twenty Years of Timing SS433}

\author{S.S. Eikenberry, P.B. Cameron, B.W. Fierce, D.M. Kull,
D.H. Dror, J.R. Houck}
\affil{Astronomy Department, Cornell University, Ithaca, NY
14853} 

\author{B. Margon}
\affil{Space Telescope Science Institute, 3700 San Martin Drive, Baltimore, MD  21218}

\begin{abstract}

	We present observations of the optical ``moving lines'' in
spectra of the Galactic relativistic jet source SS433 spread over a
twenty year baseline from 1979 to 1999.  The red/blue-shifts of the
lines reveal the apparent precession of the jet axis in SS433, and we
present a new determination of the precession parameters based on these
data.  We investigate the amplitude and nature of time- and
phase-dependent deviations from the kinematic model for the jet
precession, including an upper limit on any precessional period
derivative of $\dot P < 5 \times 10^{-5}$.  We also dicuss the
implications of these results for the origins of the relativistic jets
in SS433.

\end{abstract}


\section{Introduction}

	SS433 is the first known example of a Galactic relativistic
jet source, and thus the forerunner of modern microquasar astrophysics.
The optical spectrum of this object shows a number of strong, broad
emission lines of the Balmer and HeI series, as well as several lines
at unusual wavelengths.  These latter have been identified as
red/blue-shifted Balmer and HeI emission from collimated jets with
intrinsic velocities of $v \simeq 0.26c$ \citep{AbellMargon}.
Furthermore, the Doppler shifts of these features change with time in
a cosinusoidal manner, leading to the label of ``moving lines''.  This
behavior is now widely accepted to be a symptom of precession of the
jet axis in SS433 on a timescale of $\sim 164$ days \citep{Margon84}.

	Early studies of the precession in SS433 indicated possible
instabilities or drifts in the precessional clock \citep{Anderson},
which could give considerable insight into the accretion processes
which must provide the precessional torque.  However,
\citet{MargonAnderson} reviewed ten years of SS433 timing data and
concluded that while significant deviations from cosinusoidal behavior
exist in SS433, the evidence for systematic long-term drifts
(e.g. precessional period derivative, $\dot P$) remained inconclusive.

	In this paper, we take the data set considered by
\citet{MargonAnderson} and add to it more than 50 Doppler shift
measurements spread over 10 years, including 9 Doppler shifts measured
in 1999.  Combined, these observations span more than 20 years, and
thus provide an excellent data set for constraining long-term drifts
in SS433's precessional clock.  We discuss the observations in Section
2.  In Section 3, we present analyses of the entire data set in the
context of the ``kinematic model'' for SS433's precessing jets.  In
Section 4, we discuss the results of these analyses, and in Section 5
we present our conclusions.

\section{Observations}

	The primary observations used here are optical spectroscopic
observations of SS433, from a wide range of telescopes and instruments
(see \citet{MargonAnderson} and references therein for details).  The
net result of these observations, spread over the period from June
1978 to July 1992, is the measurement of 433 Doppler shifts for the
``receding jet'' ($z_1$) and 482 Doppler shifts for the ``approaching
jet'' ($z_2$) for the optical moving lines in SS433.

	We obtained further spectra of SS433 in July, 1999 using the
Hartung-Boothroyd Observatory (HBO) 24-inch telescope and optical CCD
spectrograph.  We used a 600 lines/mm grating and $6 \arcsec$ slit
providing a resolution of $R \sim 800$ ($6 \ {\rm \AA /pix}$).  We
present a typical spectrum in Figure 1.

	We determined the Doppler shift of each HBO spectrum using
only the moving $H \alpha$ lines, and we did so by fitting a Gaussian
profile to the red and blue components separately.  Note that the
profiles of the moving lines are broad, time-variable, and often
asymmetric.  This is due to the time overlap of multiple discrete
emission components, commonly referred to as ``bullets''
\citep{Vermeulen}, with typical lifetimes of $\sim 3$ days.  These
systematic deviations introduce a relatively large uncertainty in the
Doppler shift determination.  Based upon examination of many spectra,
we find that the typical full-width at half-maximum (FWHM) is $\Delta z
\sim 0.003$, and we adopt this as our uncertainty in the Doppler shift
determination $\sigma_z$.

\section{Analysis}

\subsection{The Kinematic Model}

	Throughout our analysis of these data, we adopt the
``kinematic model'' for the moving lines, which assumes that the
changing Doppler shifts arise from the precession of the jet axis in
SS433.  The simplest form of the kinematic model takes into account
five components: the jet velocity $\beta = {v \over{c}}$; the jet
angle from the precessional axis $\theta$, the inclination angle of
the system with respect to the observer's line of sight $i$, the
precession period $P$, and the epoch of zero precessional phase $t_0$.
The period and zero-phase epoch combine to give the precessional phase
$\phi = {{t - t_0} \over{P}}$.  The resulting Doppler shifts obey the
equation

$$ z_{1,2} = 1 - \gamma [1 \pm \ \beta \ \sin \theta \ \sin i \ \cos \phi \ \pm \ \beta \ \cos \theta \ \cos i]$$

\noindent where $\gamma = (1- \beta^2)^{-1/2}$.  SS433 exhibits
``nodding'' of the jets on a $\sim 6.5$-day period \citep{Katz} due to
the $\sim13$-day binary motion of SS433 \citep{Crampton}, which are
not accounted for in this model.  However, the effects of this nodding
are essentially negligible for long timescale studies of the jets such
as ours.  We further mitigate the impact of nodding by applying a
7-day boxcar smoothing filter to the individual Doppler shifts
determined above.  We then used chi-squared minimization to find the
best-fit parameters for the kinematic model (Table 1).  The resulting
model fit is plotted versus time along with the data and residuals in
Figures 2-3 for $z_1$ and $z_2$.  We plot the same model fit, data,
and residuals versus precessional phase in Figure 4.

	The resulting fit has a chi-squared residual per degree of
freedom of $\chi^2_{\nu} = 8.9$, indicating the presence of
statistically significant residuals.  However, we can still use this
fit to estimate uncertainties in the kinematic model parameters as
follows.  First, we scale all of the $\sigma_z$ values by
$\sqrt{8.9}$, so that the residuals have $\chi^2_{\nu-fix} = 1.0$,
essentially by fiat.  We then take the uncertainties to be the range
of a model parameter which introduces a total change of $\Delta
\chi^2_{fix} = 1.0$.  We also report these values in Table 1.  This
rescaling approach for deriving the model parameter uncertainties is
statistically valid in a strict sense only if the residuals are
consistent with Gaussian noise and are not correlated with any model
parameters in a systamtic way.  If so, then the residuals would simply
indicate that we have ignored one or more sources of noise in the
system when estimating the uncertainties in the individual Doppler
shifts.  As we show below, this is largely true, though we see some
evidence of small (but statistically-significant) systematic
deviations from the kinematic model.  Thus, the uncertainties in the
model parameters in Table 1 are likely to be good, but not perfect,
statistical estimates.  For the remainder of the paper, we adopt the
best-fit model parameters presented in Table 1.

\subsection{Doppler Shift Residuals}

	One obvious feature of Figures 2-4 is that the residuals to
the model fit greatly exceed the uncertainties in the Doppler shift
determinations (as also shown by the large value of $\chi^2_{\nu}$
above).  We also notice no obvious trend in the residuals versus time
as would be expected for systematic timing effects, such as a constant
precessional period time derivative $\dot P$.  Such large, apparently
random residuals have been noticed in previous timing studies of SS433
(\citet{Anderson}; \citet{MargonAnderson}).

\subsubsection{Correlations in Residuals}

	Previous studies have also noticed that the velocity residuals
in SS433 show a pattern of correlation between $z_1$ and $z_2$
\citep{MargonAnderson}.  Specifically, when we plot the residuals of
$z_1(obs)-z_1(mod)$ versus $z_2(obs)-z_2(mod)$, we find that most of
the points lie in the second and fourth quadrants (Figure 5).  In
other words, when the absolute value of $z_1$ is greater than expected,
the absolute vlaue of $z_2$ is also greater than expected, and vice
versa.  The number of data points with $z_1$ and $z_2$ residuals in
quadrants 2 and 4 is $271 \pm 16$, while in quadrants 1 and 3 the
number is $110 \pm 10$ -- a $>8 \sigma$ difference.

	The linear correlation coefficient between the residuals is $r
= -0.69 \pm 0.02$.  We estimate the uncertainty from a Monte Carlo
simulation as follows.  We take the 381 pairs of $z_1$ and $z_2$
residuals and add to each a random number drawn from a Gaussian
distribution with mean of zero and a standard deviation of 0.003 --
the typical uncertainty in the Doppler shift measurements.  We then
calculate the correlation coefficient of the resulting simulated
distribution.  We repeat this procedure 1000 times and then take the
standard deviation in the correlation coefficient as the uncertainty
above, $\sigma_r = \pm 0.02$.

	This correlation pattern could have several physical sources.
The effect considered most commonly in previous studies
(e.g. \citet{MargonAnderson}) is that of phase noise in the
precessional motion, with strict symmetry between $z_1$ and $z_2$.  As
the jet precessional phase either lags or leads the model ephemeris,
the projected velocity amplitudes of the jets on the observers line of
sight will either exceed or fall short of the model prediction.
Another possible physical explanation is modulation of the velocity
amplitude -- ``$\beta$-noise'' -- in a system which otherwise follows
the 5-parameter kinematic model ideally (e.g. \citet{Milgrom}).

\subsubsection{Phase-Dependence of Residuals}

	Another factor which could impact the $z_1/z_2$ residual
correlation are phase-dependent residuals to the kinematic model.
\citet{MargonAnderson} found no evidence for such phase-dependence in
their data.  In analyzing this data set, we divided the data into 10
evenly-spaced phase intervals and calculated the average and standard
deviation of the residual Doppler shifts in each bin (Table 2).  None
of the average residuals from the kinematic model is as large as the
rms deviation for residuals in that phase bin, indicating that
phase-dependent deviations from the kinematic model do not dominate
the residuals.  Furthermore, if we compare the average residuals of
the individual phase bins to the rms scatter of all the phase bins,
none of them are more than $2 \sigma$ outliers.  Thus, the amplitude
of any average deviation from the model velocity does not seem to be
phase-dependent.

	However, when we calculate the uncertainty in the average
residual for each phase bin, equal to the standard deviation in the
residuals divided by the square-root of the number of points, we find
that the deviations are in fact statistically significant (Table 2).
That is, while the scatter around the kinematic model in any given
phase bin is not dominated by systematic deviations from the model,
such deviations are present in the data set.  The nature of these
systematic deviations are not clearly determined.  Figures 2-3 show
that the velocity residuals show some correlation on timescales of
weeks or months.  Thus, sparse sampling of the velocities combined
with such correlations in the residuals could be one explanation for
the apparent systematic deviations.

	We used the average residuals from the kinematic model as
correction factors for the quadrant analysis presented above and in
Figure 5, taking the observations and subtracting both the kinematic
model and the average deviation for all points within $\Delta \phi =
\pm 0.05$ cycles of each data point.  As a result, the number of
$z_1$,$z_2$ residual pairs in quadrants 2 and 4 does decrease, but only
to 261 (with 120 pairs in quadrants 1 and 3).  Thus the correlation
between $z_1$ and $z_2$ residuals remains highly statistically
significant even after correction for the systematic deviations.

	We also note that the relative phase-independence of the
residuals raises questions regarding the nature of the residuals.  As
can be seen from the equation for Doppler shifts in the kinematic
model, every parameter which {\it could} be time-variable ($\beta, P,
i, \theta, $) either feeds directly into the phase $\phi$ or is
multiplied by $\cos \phi$.  Thus, noise in these terms or in $\phi$
itself should result in a cosinusoidal modulation in the RMS of the
Doppler shift residuals with $\phi$.

\subsubsection{The Phase Noise Model}

	As mentioned above, the most commonly-invoked physical model
for the velocity residuals in SS433 is ``phase jitter'' in the jet
precession.  Since the precession phase affects both jets similarly,
it naturally explains the correlation between the $z_1$ and $z_2$
residuals.  If such jitter can occur over timescales of weeks or
months, it can also explain the long-term residual correlations
evident in Figures 2-3.

	We analyzed this SS433 data set following the example of
\citet{MargonAnderson}, determining phase errors from the velocity
residuals above.  We simply defined the phase error to be the phase
difference between the actual phase of the observation given its epoch
and the kinematic model parameters in Table 2 and the closest model
point with the same observed velocity.  As can be seen in Figure 4,
some observed velocity amplitudes exceed the maximum model velocity
amplitude, and such points were dropped from this analysis.  We then
divided the data set into 10-day intervals and calculated the average
and standard deviation of the phase errors from all phase measurements
doing that interval (including both $z_1$ and $z_2$).  For 10-day
intervals with only 1 phase measurement we have no estimate of the
standard deviation, and thus dropped such intervals from the analysis.
We plot the resulting phase noise measurements in Figure 6.  We
repeated this same analysis using a 30-day interval for averaging,
with the results shown in Figure 7.

	We note that while there are occasional trends in the
residuals on timescales of several hundred days, no obvious trend is
apparent over the full time span in either panel of Figure 7.  The
1999 data are marginally inconsistent with zero phase residual (at the
$2.8 \sigma$ level for one of the two data points in Figure 6b).
However, it is clear that this phase residual is less than many prior
apprently secular deviations from the kinematic model in Figures 6-7.
If we assume that some period derivative is present in SS433 over the
span of our observations, these secular deviations could mask its
effects up to $\Delta \phi \sim 0.05$ cycles.  Given the span of our
observations, this corresponds to an upper limit on the period
derivative of $\dot P < 5 \times 10^{-5}$.

\subsubsection{The Velocity Amplitude Noise Model}

	As mentioned above, an alternate physical explanation for the
velocity residuals in Figures 2-4 is noise in the intrinsic velocity
of the jets.  To investigate this possibility further, we calculated
the intrinsic jet velocity necessary to match each observed Doppler
shift, given $\theta, \ i , \ t_0 , \ {\rm and} \ p$ from the best-fit
parameter set in Table 1.  We plot the corresponding values for $\beta
= {v \over{c}}$ versus time in Figure 8 and versus precessional phase
in Figure 9.

	The average value of $\beta$ we find is 0.254 with a standard
deviation of $0.024$.  Given 507 independent measurements of $\beta$
in this way, we arrive at an average value of $\beta_{ave} = 0.254 \pm
0.0011$.  While this value is only $\sim 4 \%$ lower than the value
given in Table 1, the difference is statistically significant at the
$7.9 \sigma$ level.  This may indicate that ``noise'' in the Doppler
shifts may in fact be impacting the parameter estimates for the
kinematic model in a systematic way, as discussed in Section 3.1.

\section{Discussion}

\subsection{Phase Noise}

	As noted above, the upper limit on precessional period
derivative of $\dot P < 5 \times 10^{-5}$ shows that there is no large
long-term drift in the precessional timing properties of SS433.  The
presence of jitter in the system implies some ``torque noise'' in the
process driving the precession, according to the phase noise model.
However, if this were the case, that noise must average out over
timescales of $\sim 20$ years.  We can also see from Figure 7 that
there are fairly large phase deviations of $\Delta \phi \sim 0.1$
cycles over timescales as short as $\sim 10$ days.  This implies that
the torque noise $\Delta \tau$ has a maximum relative amplitude of at
least

$$ {\Delta \tau_{max} \over{\tau}} \simeq {{(\Delta \phi / \Delta t)} \over{(\partial \phi / \partial t)}}$$

$$	\simeq 1.6$$

\noindent Thus, the variation in torque can in fact exceed the
time-averaged torque driving the precession.  This may be a problem
for certain physical models of the precession and timing noise in
SS433.

	Finally, we note that the phase noise model is incapable of
producing the observed Doppler shifts which exceed the maximum
amplitude predicted by the kinematic model.  We have considered the
possibility that the phase noise itself causes the $\chi$-squared
fitting procedure used to determine the model parameters to
systematically underestimate the true jet velocity, and thus
``undershoot'' the maxima.  However, Monte Carlo simulations of data
sets with higher true velocities and phase noise identical to that
observed here fail to produce such undershooting.  Therefore, we
conclude that phase noise model cannot reproduce the observed Doppler
shift residuals near the maximum projected velocities.

\subsection{Velocity Noise}

	The alternate ``$\beta$-noise'' model, on the other hand, can
clearly explain the excess velocity at the extrema (and any other
precessional phase) by the changing jet velocity amplitude.  Such a
model also has a physical basis, given recent advances in the modeling
of relativistic jet production.  \citet{meier} dicuss a scenario where
such jets are launched by a magnetic accretion disk instability around
a black hole (or other compact object).  Variations in the accretion
flow onto the compact object (i.e. $\dot M$, intrinsic magnetic field,
etc.) can alter the radius at which the magnetic field saturates and
the jet is launched, and thus the jet velocity.  The relation between
jet velocity and launch radius for a non-rotating black hole follows:

$$ \beta (R) = \sqrt{{2 R_g \over{R}}} $$

\noindent where $\beta = {v \over{c}}$, and $R_g$ is the gravitational
radius of the black hole (one-half of the Schwarzschild radius).

\subsection{Jitter models and phase-dependence of residuals}

	It is also interesting to view these model in light of the
apparent lack of phase-dependence in the Doppler shift residuals noted
above.  By differentiating the equation for Doppler shifts in the
kinematic model (eqn. 1) with respect to the potentially time-varying
model components ($\beta$, $i$, $\theta$, and $\phi$) we can see the
relative phase-dependence of Doppler shift residuals on deviations in
each term.  In the phase noise model, we would expect the following
dependence:

$$ \Delta z = (\gamma \ \beta \ \sin \theta \ \sin i \ \sin \phi) \ \Delta \phi $$
$$	\simeq 0.1 \sin \phi \ \Delta \phi $$

\noindent Thus, we would expect the amplitude of the Doppler shift
residuals to be sinusoidally modulated with phase, in apparent
contradiction with our analyses above.

For the ``$\beta$-noise'' model, we have:

$$ \Delta z = (\gamma \ \sin \theta \ \sin i \ \cos \phi \ - \ \gamma \ \cos \theta \ \cos i) \ \Delta \beta $$
$$	\simeq (0.35 \cos \phi \ - \ 0.2) \ \Delta \beta $$

\noindent (with the approximation that $\partial \gamma / \partial
\beta \simeq 0$).  Again, we have a modulation of the Doppler shift
residual amplitude dominated by a term varying cosinusoidally with
respect to phase.

	For variations in the angle between the jet axis and the
precessional axis, $\theta$, we have:

$$ \Delta z = (\gamma \ \beta \ \cos \theta \ \sin i \ \cos \phi \ + \ \gamma \ \beta \ \sin \theta \ \cos i) \ \Delta \theta $$
$$	\simeq (0.24 \cos \phi \ + \ 0.09) \ \Delta {\theta} $$

\noindent again dominated by a $\cos \phi$ term.

	Finally, for variations in the system inclination angle, $i$,
we  have:

$$ \Delta z = (\gamma \ \beta \ \sin \theta \ \cos i \ \cos \phi \ + \ \gamma \ \beta \ \cos \theta \ \sin i) \ \Delta i $$
$$	\simeq (0.02 \cos \phi \ + \ 0.24) \ \Delta i $$

\noindent Interestingly, the Doppler shift residual amplitudes for
``$i$-noise'' would be dominated by a constant term, with only a small
dependence on phase.  Thus, this is the only parameter in the
kinematic model for which variations causing the Doppler shift
residuals are consistent with their observed phase-independence.
Unfortunately, we know of no physical model for such variability at
this time.  Furthermore, based on the mass estimates for the compact
object and companion star ($\sim 10 M_{\odot}$ total), the known
13.5-day binary period, and the presence of deviations as large as
0.01 rad/day in inclination angle, the change in rotational energy
would require average powers of $\dot E \sim 10^3 \ L_{Edd}$, which
seems implausible.

	One other possible explanation is that the Doppler shift
residuals are due to variations in both precessional phase and one (or
more) of the other model parameters.  In that case, the residual
amplitude would depend on a sum of $\cos \phi$ and $\sin \phi$ terms
which could potentially smooth out any phase-dependence in the
residuals.

\section{Conclusions}

	We have presented observations of the Doppler-shifted optical
moving lines in SS433 spanning over 20 years.  We draw the following
conclusions based on the data:

\begin{itemize}

\item We find parameters for the kinematic model for the jet
precession which are similar to those found by previous authors
(e.g. \citet{MargonAnderson}).

\item We find a strong correlation between residuals to the models
fits for the two jets $z_1$ and $z_2$, with a linear correlation
coefficient of $r = -0.69 \pm 0.02$.

\item We find that the residuals to the kinematic model fit are {\it
not} dominated by systematic phase-dependent deviations from the
model.  However, systematic phase-dependent deviations from the
kinematic model {\it are} seen in the data set at a low level.

\item If we adopt a ``phase noise'' model for the velocity residuals,
we find correlated deviations over timescales of months to years, but
no long-term trend over the full data set.  We place a limit on the
precessional period derivative of $\dot P < 5 \times 10^{-5}$.

\item Noise in any single parameter of the kinematic model seems
unable to explain the observed phase-independence of the velocity
residuals in SS433.  However, variations in both phase and one of the
other parameters would vary as the weighted sum of $\cos \phi$ and
$\sin \phi$ terms, which could smooth out any phase-dependence of the
Doppler shift residuals.

\end{itemize}

\acknowledgments The authors thank D. Lai and I. Wasserman for helpful
discussions of these results.  SSE is supported in part at Cornell by
an NSF CAREER award (NSF-9983830), and PBC was partially supported by
this grant.  DMK was supported at Cornell by a NASA Space Grant summer
research fellowship.

\vfill \eject

\begin{deluxetable}{lccccc}
\tablecaption{Best-fit parameters for the kinematic jet precession model}
\startdata
Parameter & $\beta$ & $\theta$ & $i$ & $P$ & $t_0$ \\
  & & (deg) & (deg) & (days) & (TJD) \\
\hline
Value & 0.2647 & 20.92 & 78.05 & 162.375 & 3563.23 \\
Uncertainty & $\pm 0.0008$ & $\pm 0.08$ & $\pm 0.05$ & $\pm 0.011$ & $\pm 0.11$ \\
\enddata
\end{deluxetable}

\begin{deluxetable}{lcccccc}
\tablecaption{Average residuals in Doppler shift versus phase}
\startdata
Phase Interval & $<\Delta z_1>$ & $\Delta z_1$ & $\Delta z_1$ & $<\Delta z_2>$ & $\Delta z_2$ & $\Delta z_2$\\
(cycles) & & RMS & Uncert. & & RMS & Uncert. \\
\hline
0.0-0.1 & -0.0035 & 0.0074 & 0.0010 & 0.0040 & 0.0067 & 0.0008 \\
0.1-0.2 & 0.0017 & 0.0082 & 0.0011 & -0.0024 & 0.0083 & 0.0010 \\
0.2-0.3 & -0.0005 & 0.0101 & 0.0014 & -0.0016 & 0.0093 & 0.0014 \\
0.3-0.4 & 0.0035 & 0.0076 & 0.0011 & 0.0003 & 0.0028 & 0.0004 \\
0.4-0.5 & -0.0011 & 0.0130 & 0.0032 & -0.0008 & 0.0067 & 0.0013 \\
0.5-0.6 & -0.0052 & 0.0116 & 0.0016 & 0.0031 & 0.0083 & 0.0010 \\
0.6-0.7 & 0.0039 & 0.0152 & 0.0028 & -0.0008 & 0.0135 & 0.0025 \\
0.7-0.8 & 0.0015 & 0.0092 & 0.0013 & -0.0001 & 0.0077 & 0.0010 \\
0.8-0.9 & 0.0026 & 0.0104 & 0.0013 & -0.0018 & 0.0079 & 0.0010 \\
0.9-1.0 & -0.0035 & 0.0088 & 0.0013 & 0.0007 & 0.0063 & 0.0008 \\
\enddata
\end{deluxetable}

\begin{figure}
\plotone{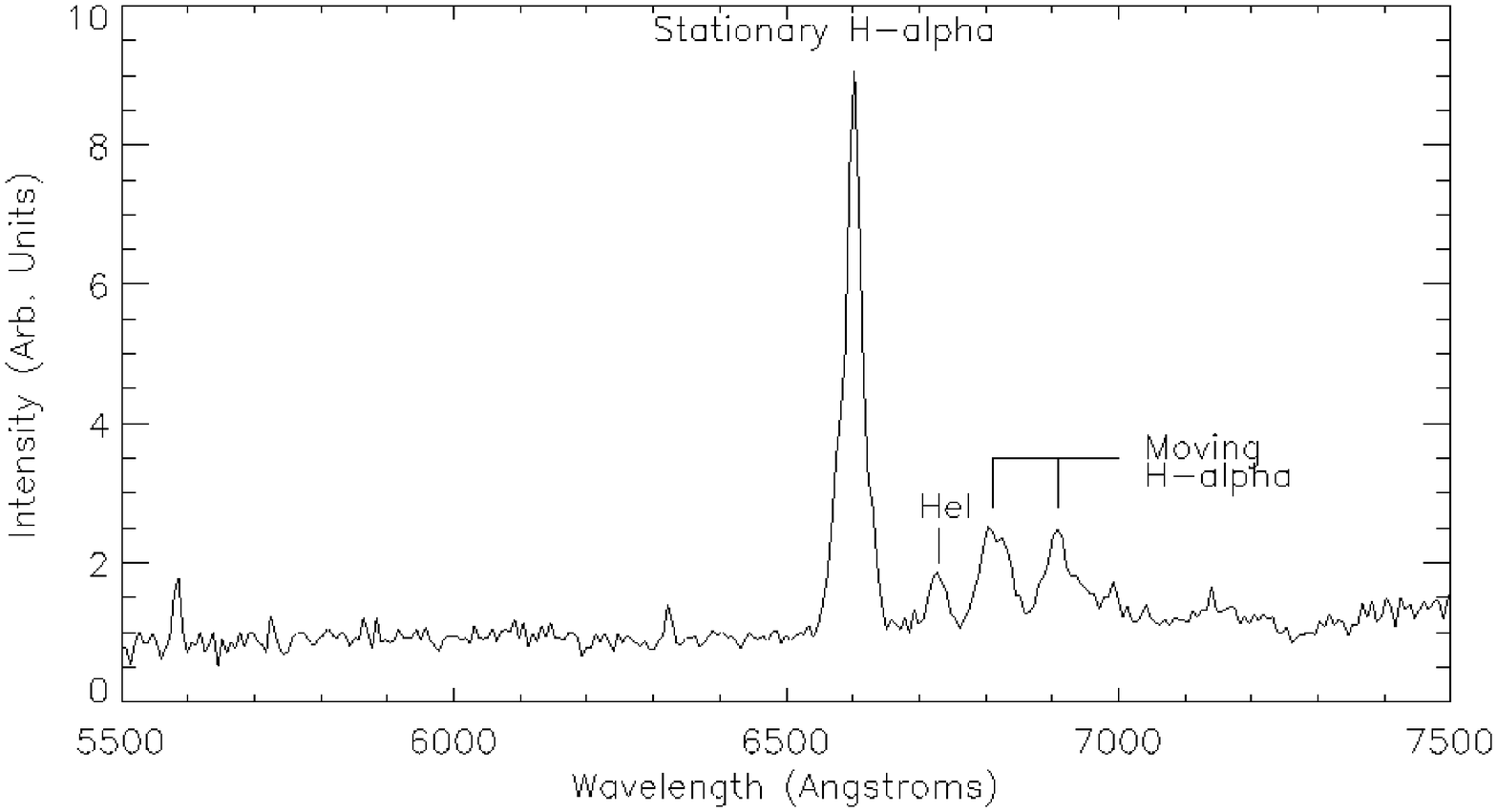}
\caption{\it Typical spectrum of SS433 taken from the
Hartung-Boothroyd Observatory 24-inch telescope.  Note the clear
appearance of the moving Balmer lines from the jets.}
\end{figure}

\begin{figure}
\epsscale{0.8}
\plotone{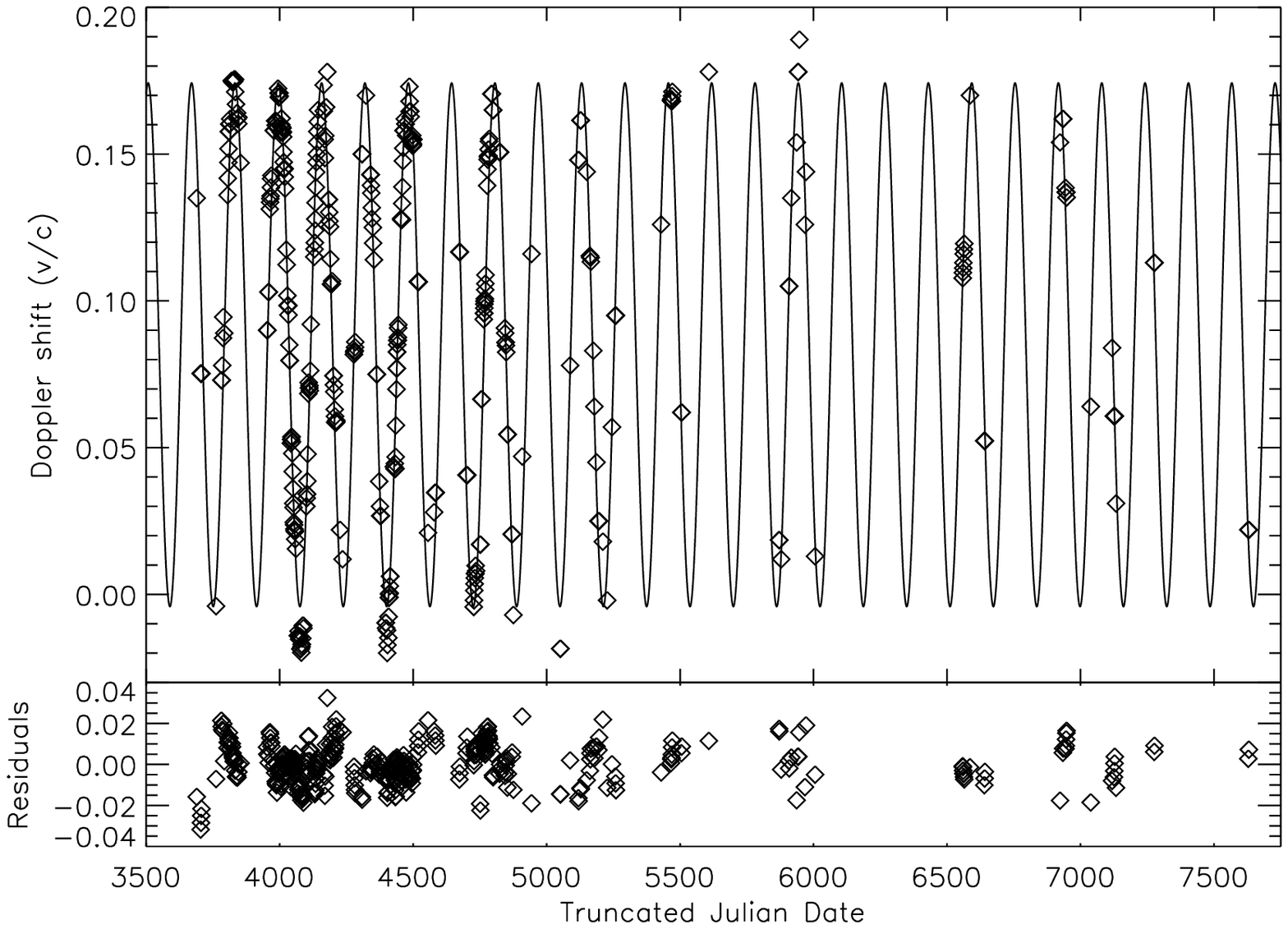}
\plotone{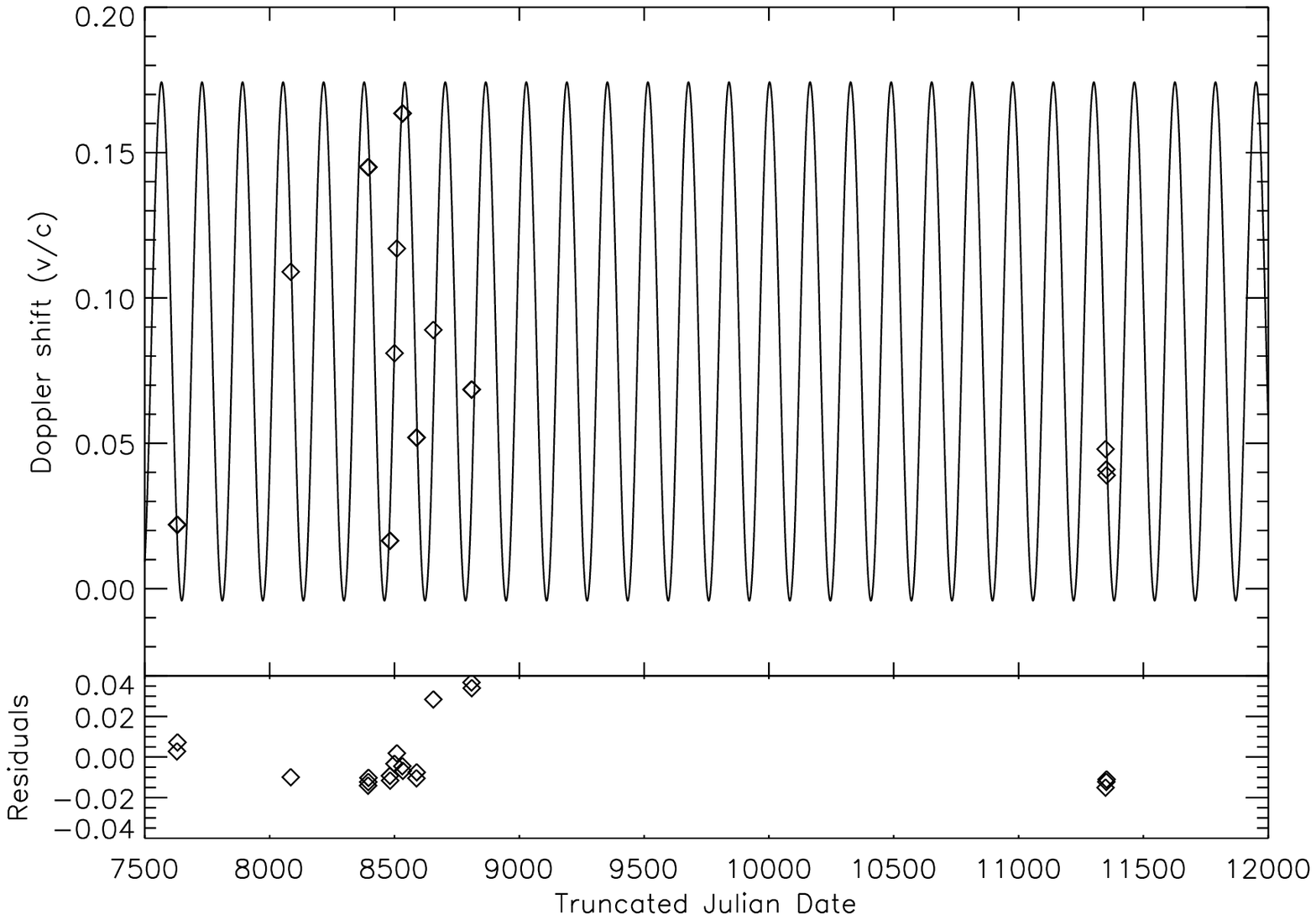}
\caption{\it Doppler shift data, model fit, and residuals versus time
based on the best-fit parameters in Table 1 for $z_1$, the ``receding
jet''.  The vertical extent of the plotting symbols shows the typical
$\pm 1 \sigma$ uncertainty in the Doppler shift.}
\end{figure}

\begin{figure}
\epsscale{0.8}
\plotone{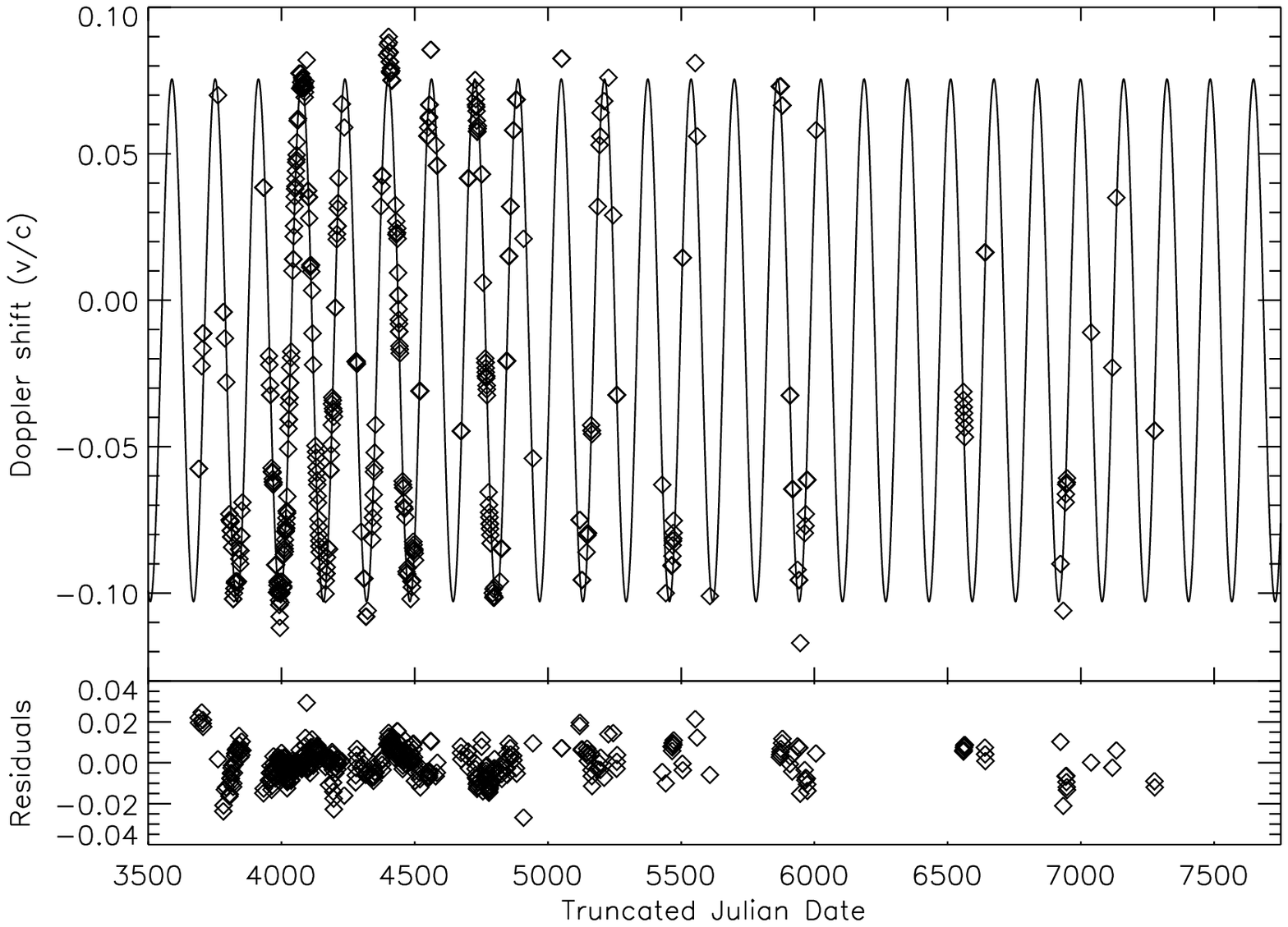}
\plotone{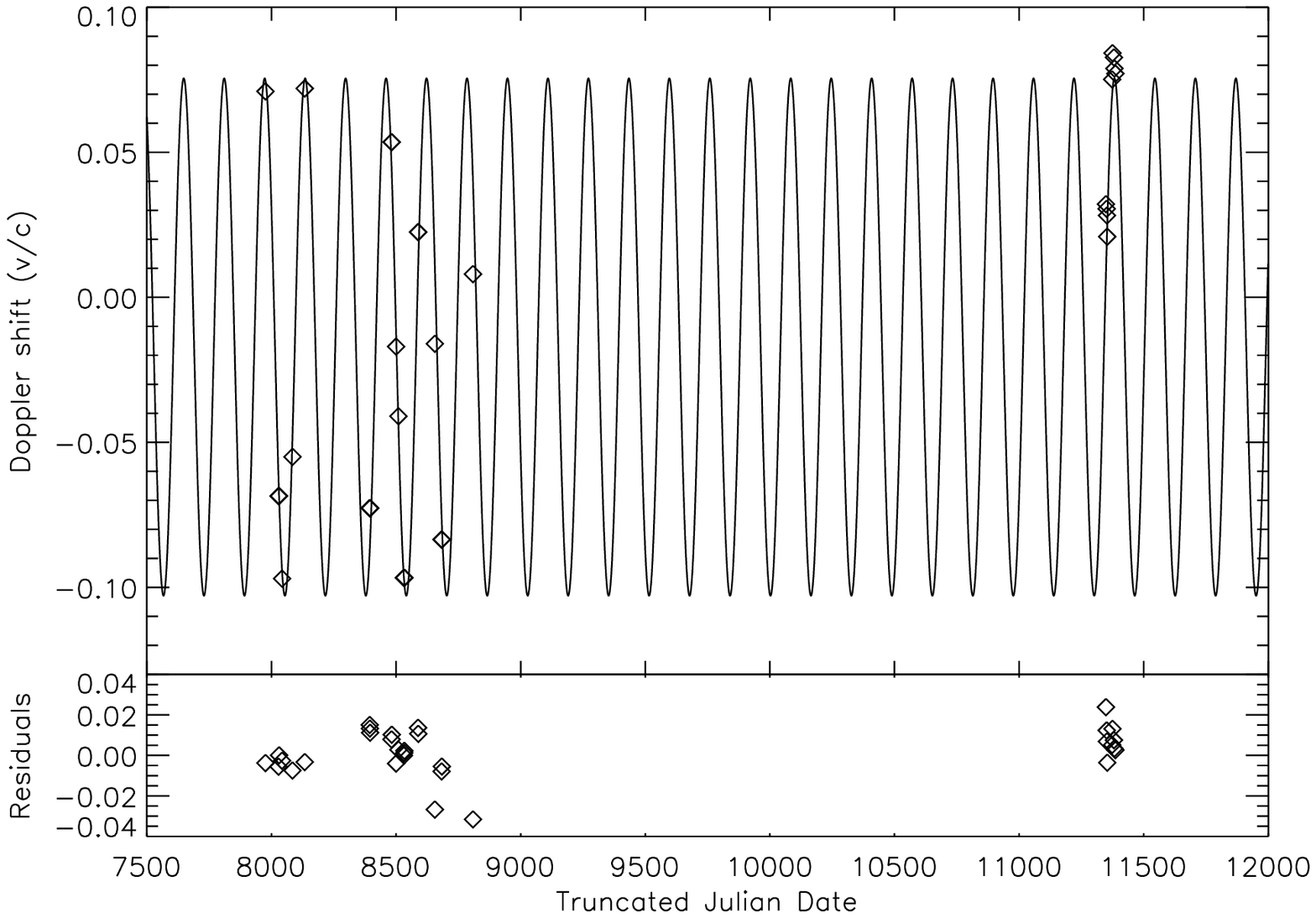}
\caption{\it Doppler shift data, model fit, and residuals versus time
based on the best-fit parameters in Table 1 for $z_2$, the
``approaching jet''.  The vertical extent of the plotting symbols
shows the typical $\pm 1 \sigma$ uncertainty in the Doppler shift.}
\end{figure}

\begin{figure}
\epsscale{0.8}
\plotone{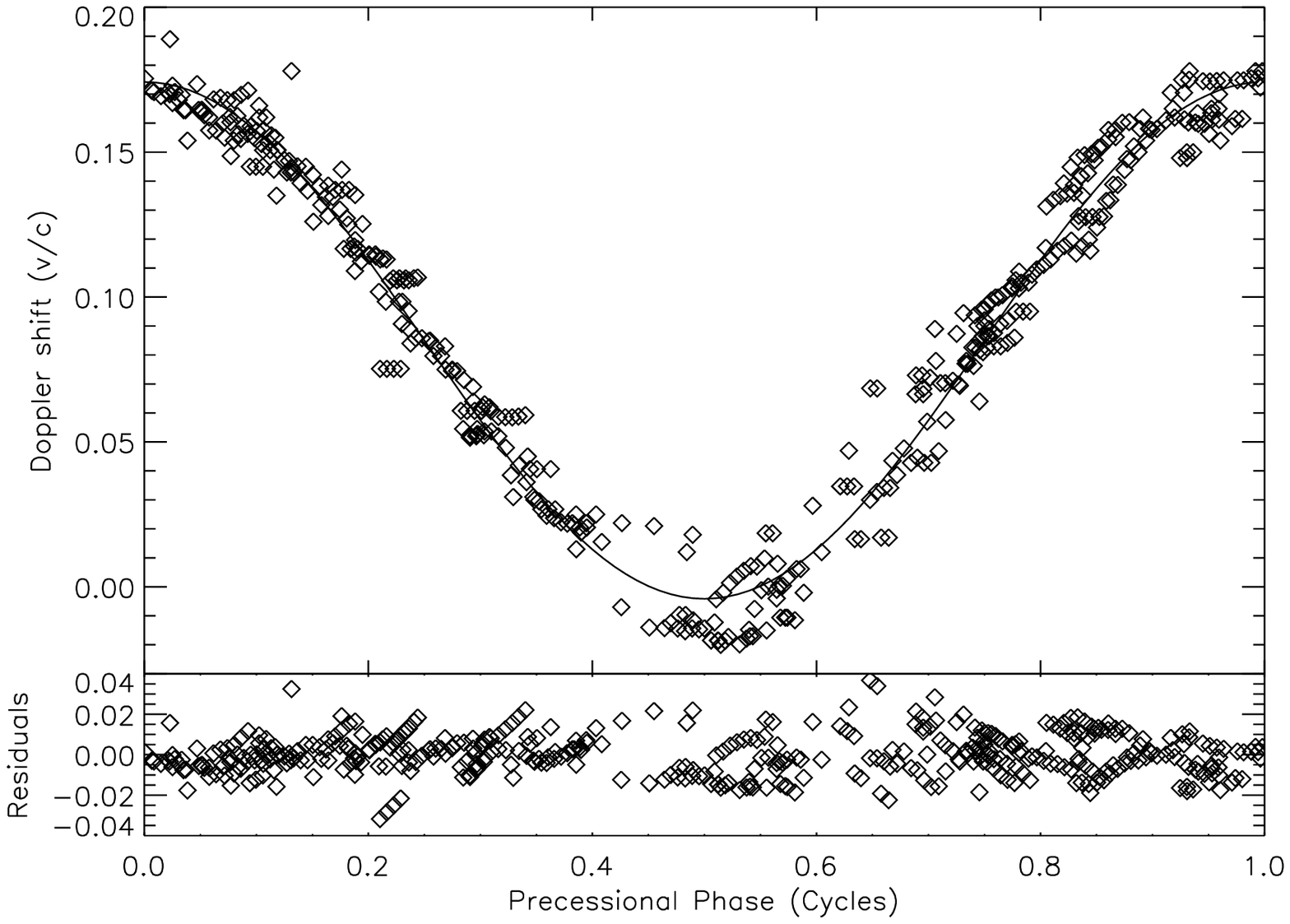}
\plotone{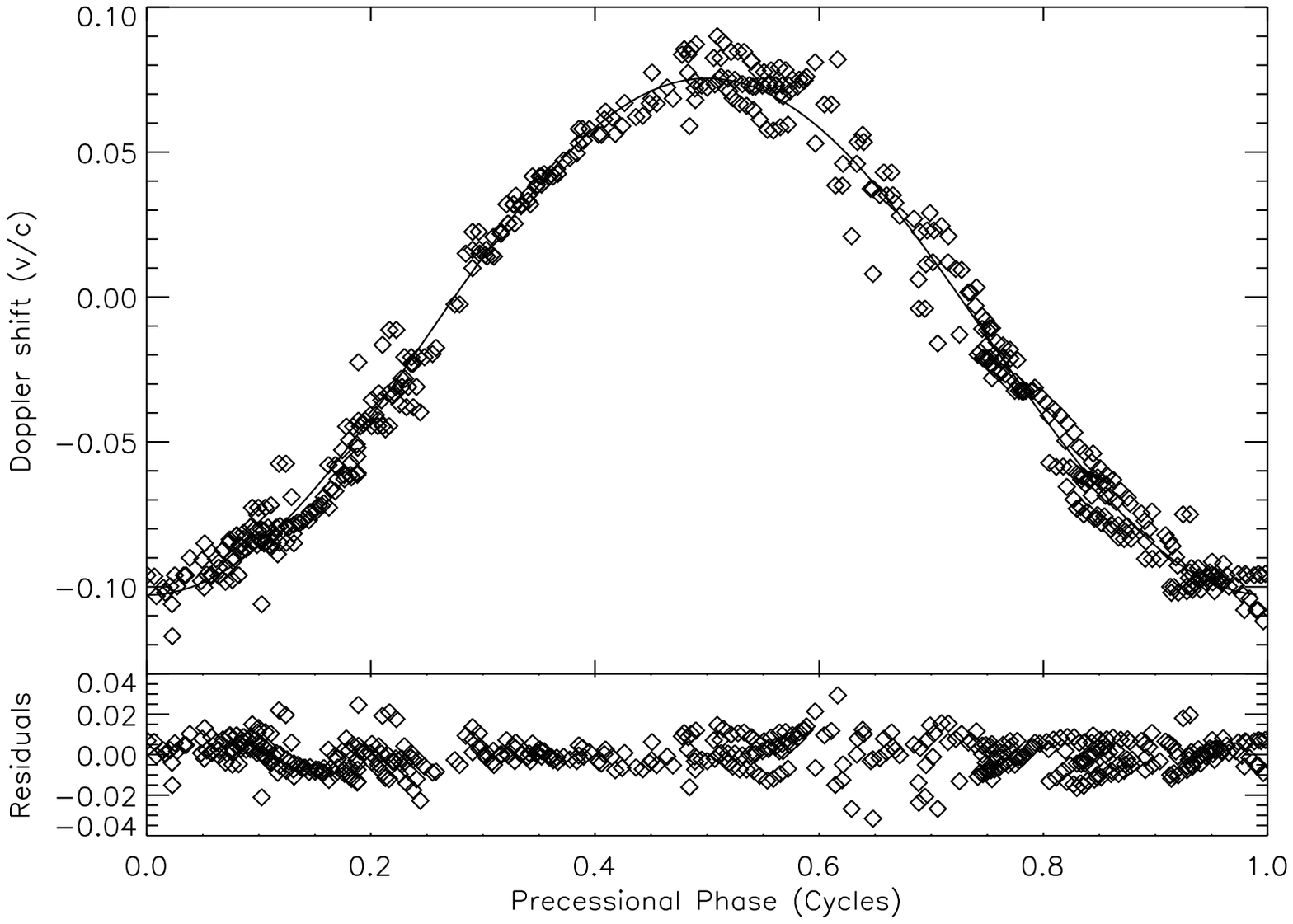}
\caption{\it Doppler shift data, model fit, and residuals versus
precessional phase based on the best-fit parameters in Table 1 for:
(top) $z_1$, the ``receding jet''; (bottom) $z_2$, the ``approaching''
jet.  The vertical extent of the plotting symbols shows the typical
$\pm 1 \sigma$ uncertainty in the Doppler shifts.}
\end{figure}

\begin{figure}
\plotone{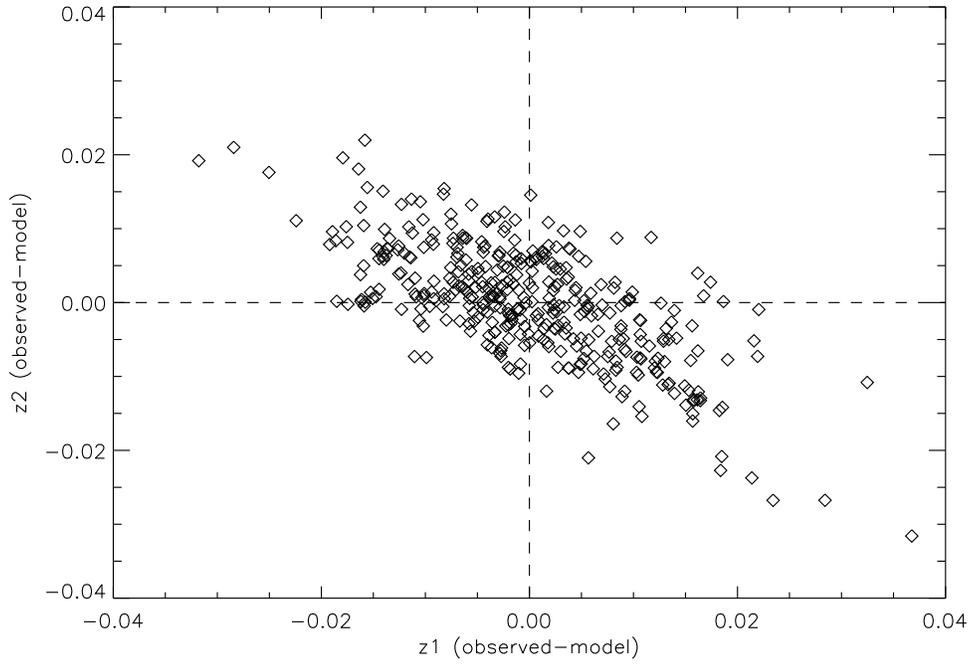}
\caption{\it Doppler shift residuals from the kinematic model.  Note the clear anti-correlation between $z_1$ and $z_2$ residuals.}
\end{figure}

\begin{figure}
\epsscale{0.8}
\plotone{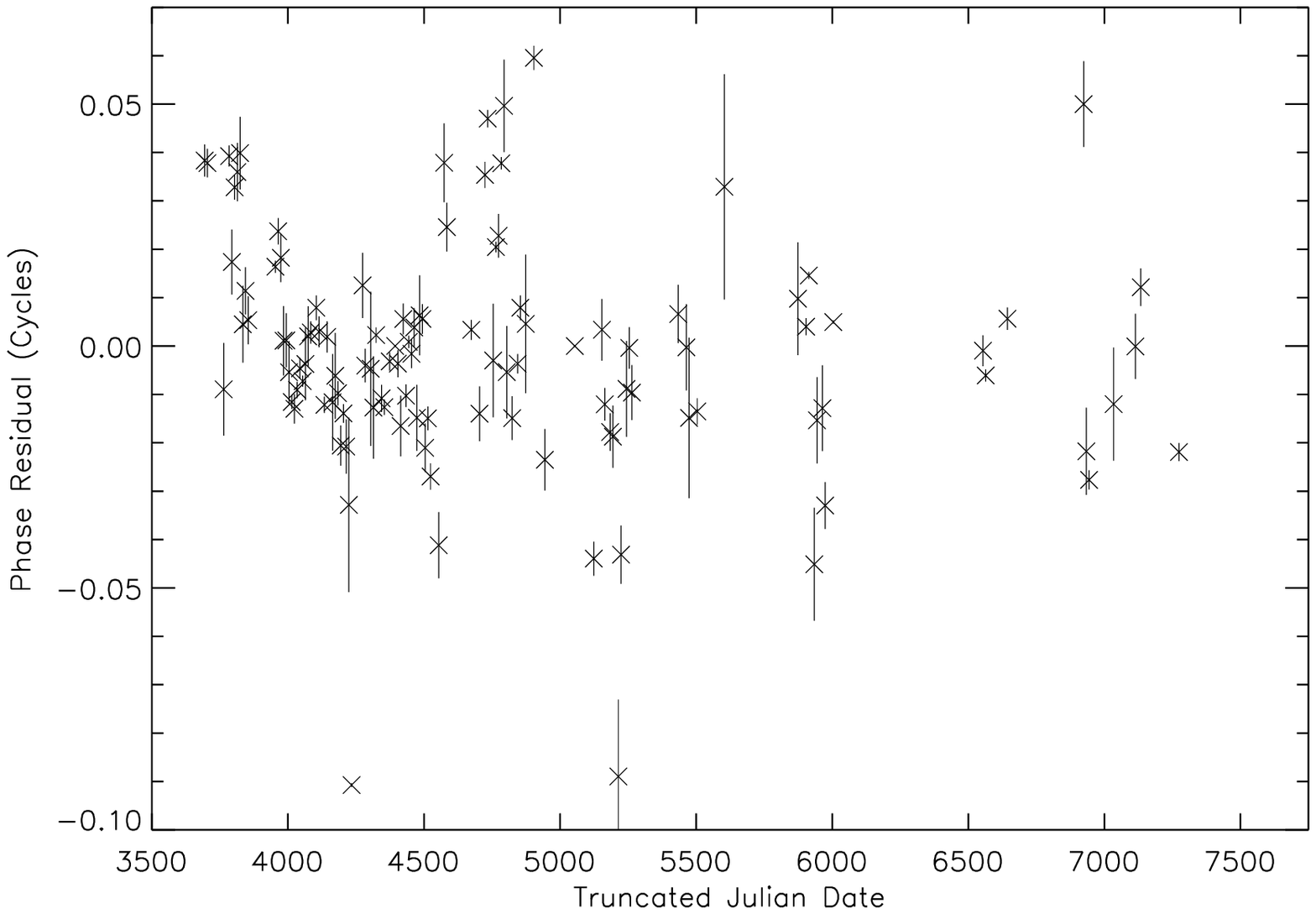}
\plotone{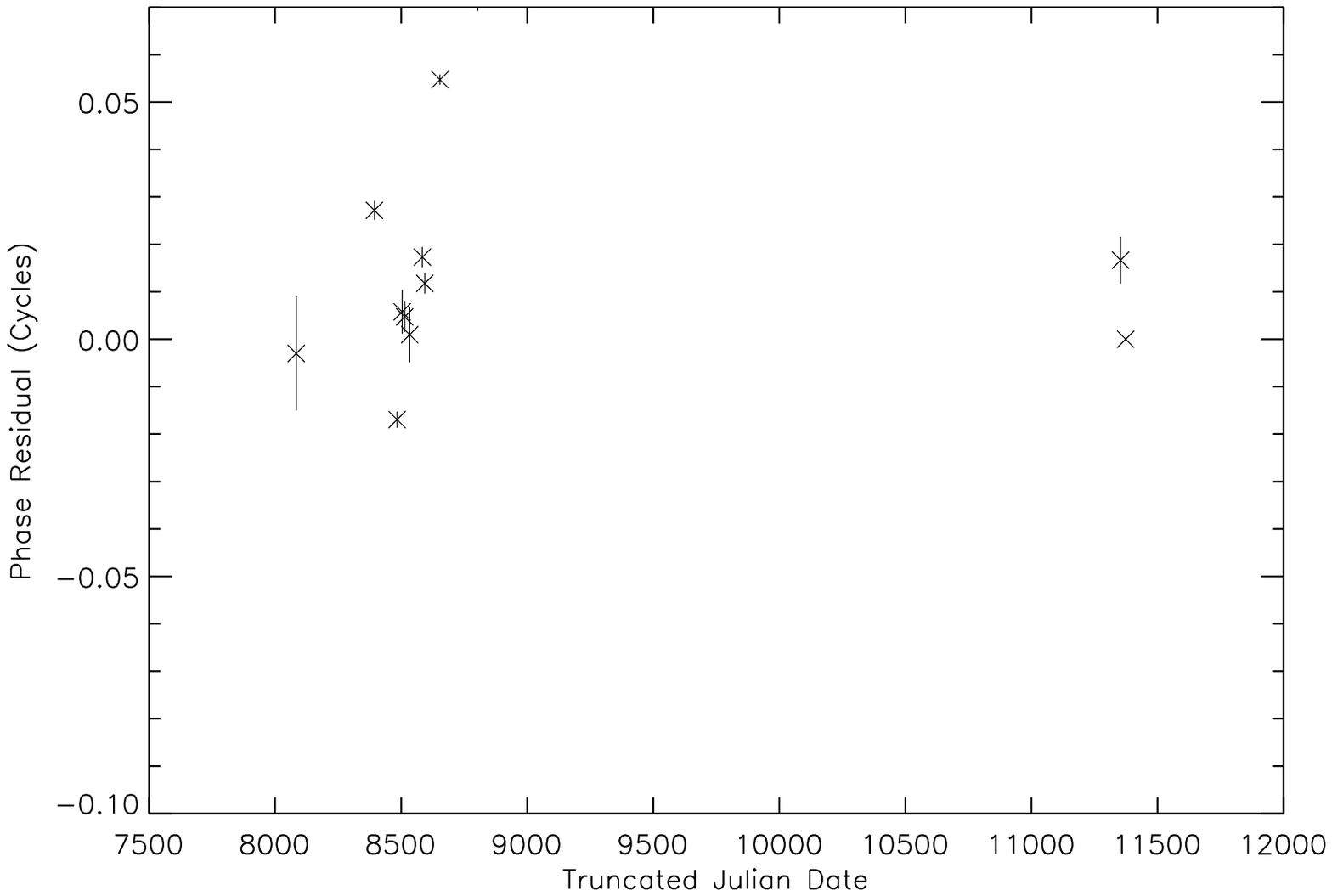}
\caption{\it Phase residuals deduced from velocity residuals to the kinematic model, averaged over 10-day intervals.}
\end{figure}

\begin{figure}
\epsscale{0.8}
\plotone{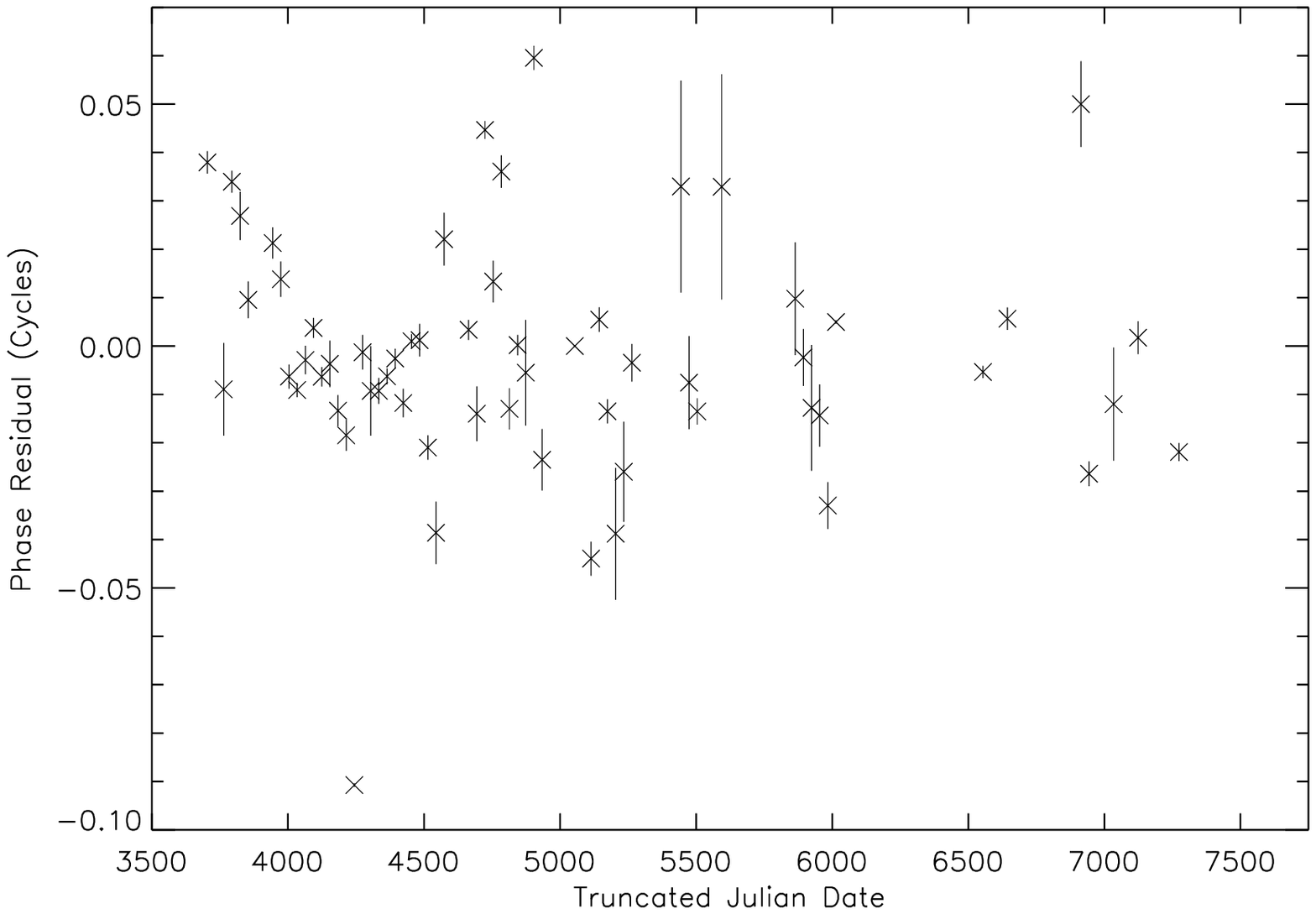}
\plotone{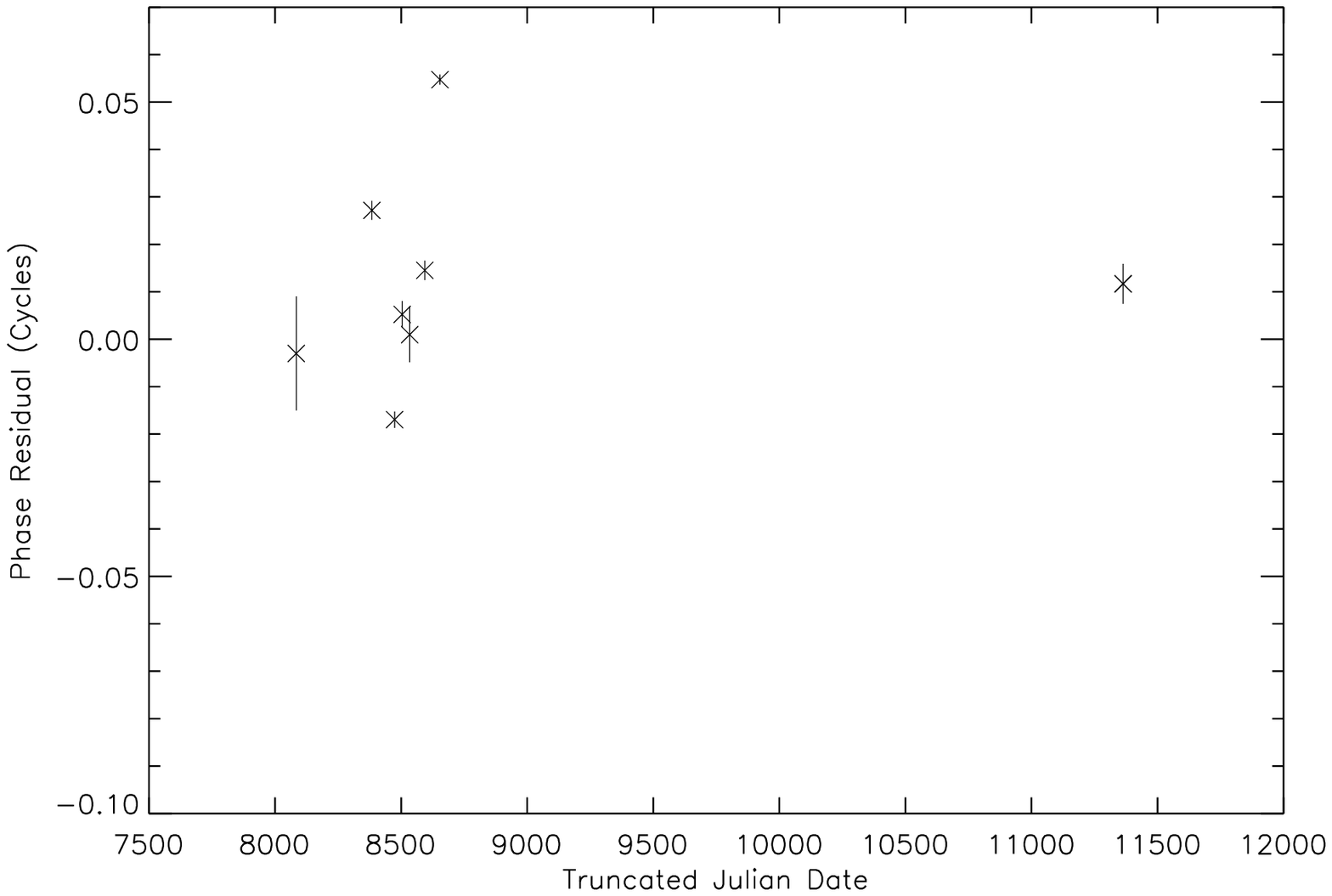}
\caption{\it Phase residuals deduced from velocity residuals to the kinematic model, averaged over 30-day intervals.}
\end{figure}

\begin{figure}
\epsscale{0.8}
\plotone{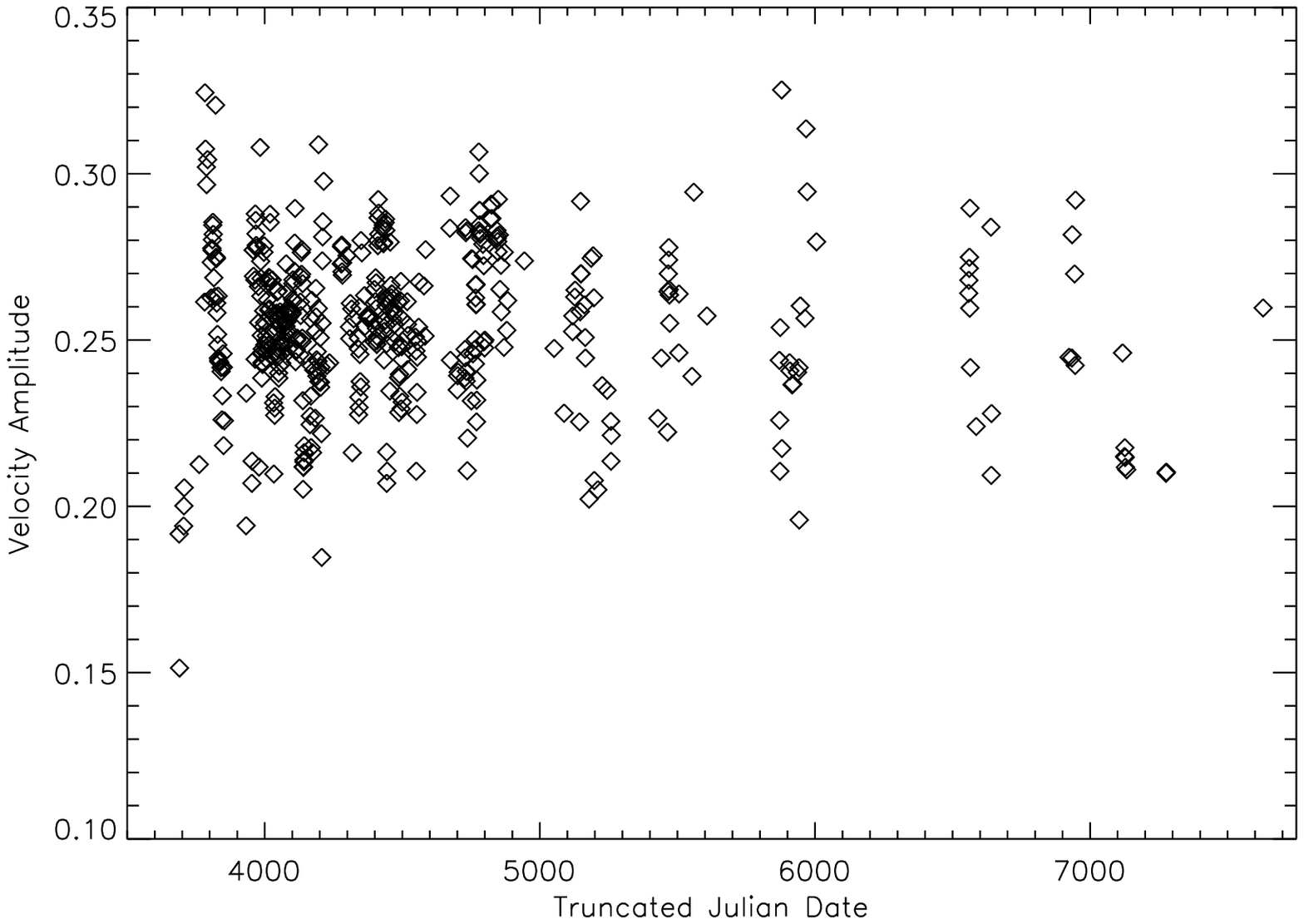}
\plotone{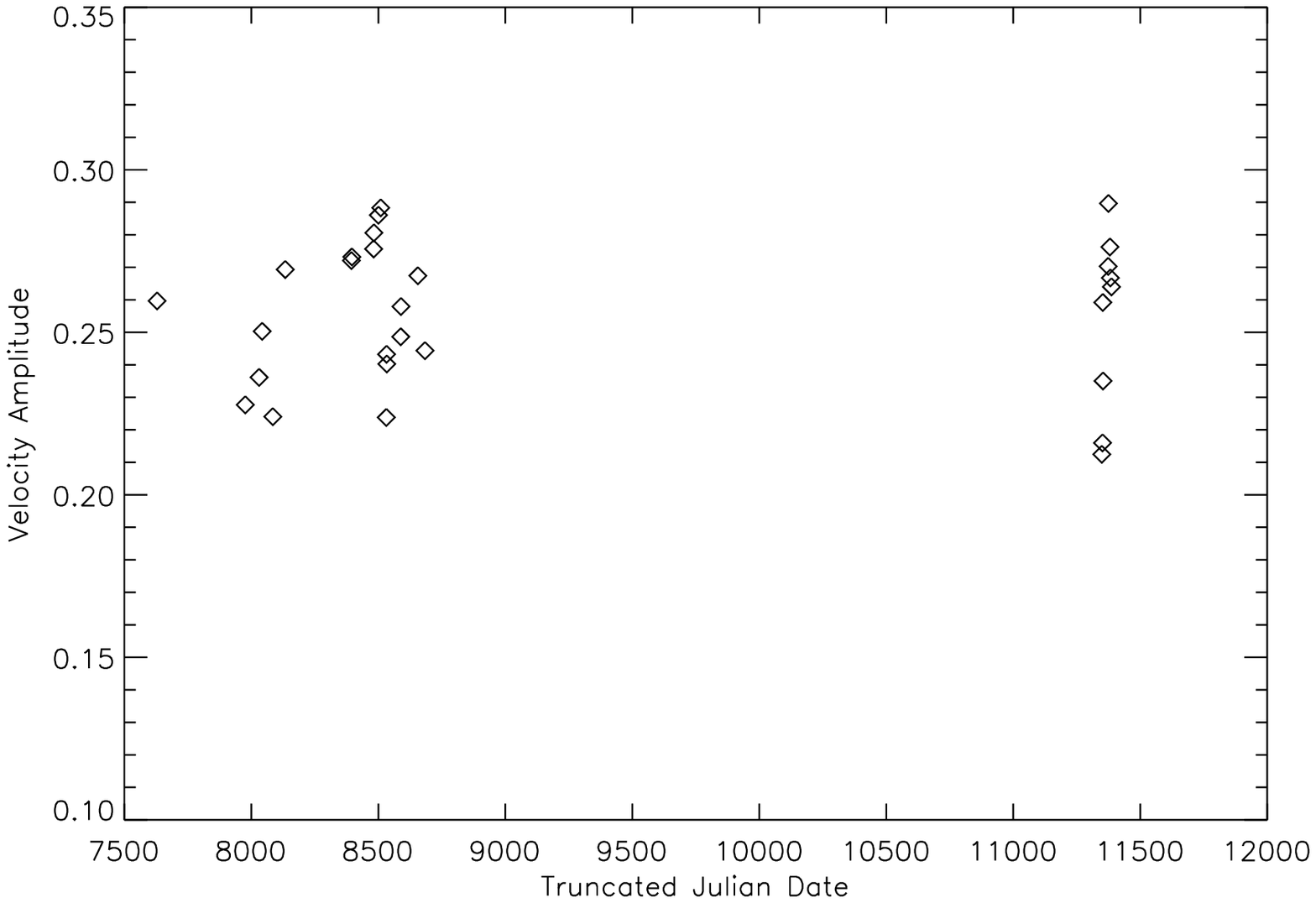}
\caption{\it Jet velocity amplitude required to match the observed
Doppler shifts as a function of time.}
\end{figure}

\begin{figure}
\plotone{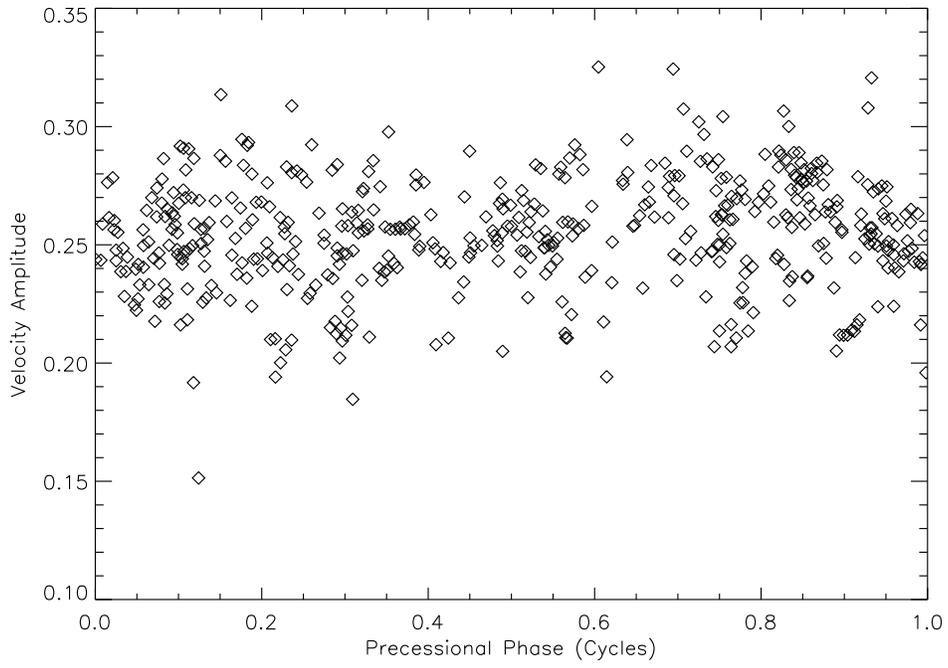}
\caption{\it Jet velocity amplitude required to match the observed
Doppler shifts as a function of precessional phase.}
\end{figure}

\end{document}